# Global Grids and Software Toolkits:
# A Study of Four Grid Middleware Technologies

Parvin Asadzadeh, Rajkumar Buyya[1], Chun Ling Kei, Deepa Nayar, and Srikumar Venugopal

Grid Computing and Distributed Systems (GRIDS) Laboratory
Department of Computer Science and Software Engineering
The University of Melbourne, Australia

## Abstract

Grid is an infrastructure that involves the integrated and collaborative use of computers, networks, databases and scientific instruments owned and managed by multiple organizations. Grid applications often involve large amounts of data and/or computing resources that require secure resource sharing across organizational boundaries. This makes Grid application management and deployment a complex undertaking. Grid middlewares provide users with seamless computing ability and uniform access to resources in the heterogeneous Grid environment. Several software toolkits and systems have been developed, most of which are results of academic research projects, all over the world. This chapter will focus on four of these middlewares—UNICORE, Globus, Legion and Gridbus. It also presents our implementation of a resource broker for UNICORE as this functionality was not supported in it. A comparison of these systems on the basis of the architecture, implementation model and several other features is included.

## 1. Introduction

The last decade has seen a substantial increase in commodity computer (PCs) and network performance, mainly as a result of faster hardware and more sophisticated software. These commodity technologies have been used to develop low-cost high-performance computing systems, popularly called clusters, to solve resource-intensive problems in a number of application domains [1]. However, there are number of problems, in the fields of science, engineering, and business, which are not tractable using the current generation of high-performance computers. In fact, due to their size and complexity, these problems are often resource (computational and data) intensive and they also need to work collaboratively with distributed interdisciplinary application models and components. Consequently, such applications require a variety of resources that are not available in a single organisation.

The ubiquity of the Internet and Web as well as the availability of powerful computers and high-speed wide-area networking technologies as low-cost commodity components is rapidly changing the computing landscape and society. These technology opportunities have prompted the possibility of harnessing wide-area distributed resources for solving large-scale problems, leading to what is popularly known as Grid computing [2]. The term "Grid" is chosen as an analogy to the electrical power grid that provides consistent, pervasive, dependable, transparent access to electric power irrespective of its source. The level of analogy that exists between electrical and computational power grids is discussed in [3].

Grids enable the sharing, exchange, discovery, selection, and aggregation of geographically/Internet-wide distributed heterogeneous resources—such as computers, databases, visualization devices, and scientific instruments. Accordingly, they have been proposed as the next-generation computing platform and global cyber-infrastructure for solving large-scale problems in science, engineering, and business. Unlike traditional parallel and distributed systems, Grids address issues such as security, uniform access, dynamic discovery, dynamic aggregation, and quality-of-services. A number of prototype applications have been developed and scheduling experiments have been carried out within grids [4] - [8]. The results of these efforts demonstrate that the Grid computing paradigm holds much promise. Furthermore, Grids have the potential to allow the sharing of scientific instruments such as particle accelerator (CERN Large Hadron Collider [9]), Australian radio telescope [10] and synchrotron [11] that have been commissioned as national/international infrastructure due to the high cost of ownership and to support on-demand and real-time processing and analysis of data generated by them. Such a capability will radically enhance the possibilities for scientific and technological research and innovation, industrial and business management,

---

[1] Contact author email id – raj@cs.mu.oz.au

application software service delivery and commercial activities, and so on.

A high-level view of activities involved within a seamless, integrated computational and collaborative Grid environment is shown in Figure 1. The end users interact with the Grid resource broker that performs resource discovery, scheduling, and the processing of application jobs on the distributed Grid resources. In order to provide users with a seamless computing environment, the Grid middleware systems need to solve several challenges originating from the inherent features of the Grid [12]. One of the main challenges is the heterogeneity in grid environments, which results from the multiplicity of heterogeneous resources and the vast range of technologies encompassed by the Grid. Another challenge involves the multiple administrative domains and autonomy issues because of geographically distributed grid resources across multiple administrative domains and owned by different organizations. Other challenges include scalability (problem of performance degradation as the size of Grids increases) and dynamicity/ adaptability (problem of resource failing is high). Middleware systems must tailor their behavior dynamically and use the available resources and services efficiently and effectively.

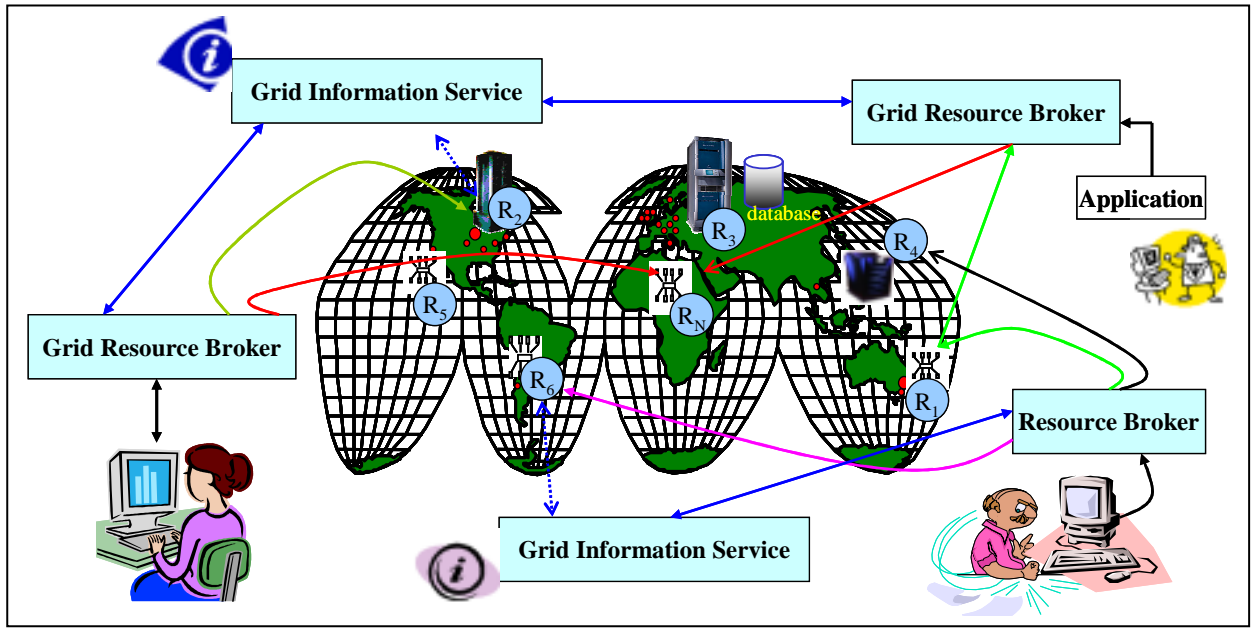

**Figure 1: A world-wide Grid computing environment.**

A tremendous amount of effort has been spent in the design and implementation of middleware software for enabling computational Grids. Several of these software packages have been successfully deployed and it is now possible to build Grids beyond the boundaries of a single local area network. Examples of Grid middleware are UNICORE (UNiform Interface to COmputing REsources) [13], Globus [13], Legion [16] and Gridbus [17]. These middleware systems aim to provide a grid-computing infrastructure where users link to computer resources, without knowing where the computing cycles are generated.

The remainder of this chapter provides an insight into the different Grid middleware systems existing today, followed by the comparison of these systems, and also casts some light on the different projects using the abovementioned middleware.

## 2. Overview of Grid Middleware Systems

Figure 2 shows the hardware and software stack within a typical Grid architecture. It consists of four layers: fabric, core middleware, user-level middleware, and applications and portals layers.

The *Grid Fabric* level consists of distributed resources such as computers, networks, storage devices and scientific instruments. The computational resources represent multiple architectures such as clusters, supercomputers, servers and ordinary PCs which run a variety of operating systems (such as UNIX variants or Windows). Scientific instruments such as telescope and sensor networks provide real-time data that can be transmitted directly to computational sites or are stored in a database.

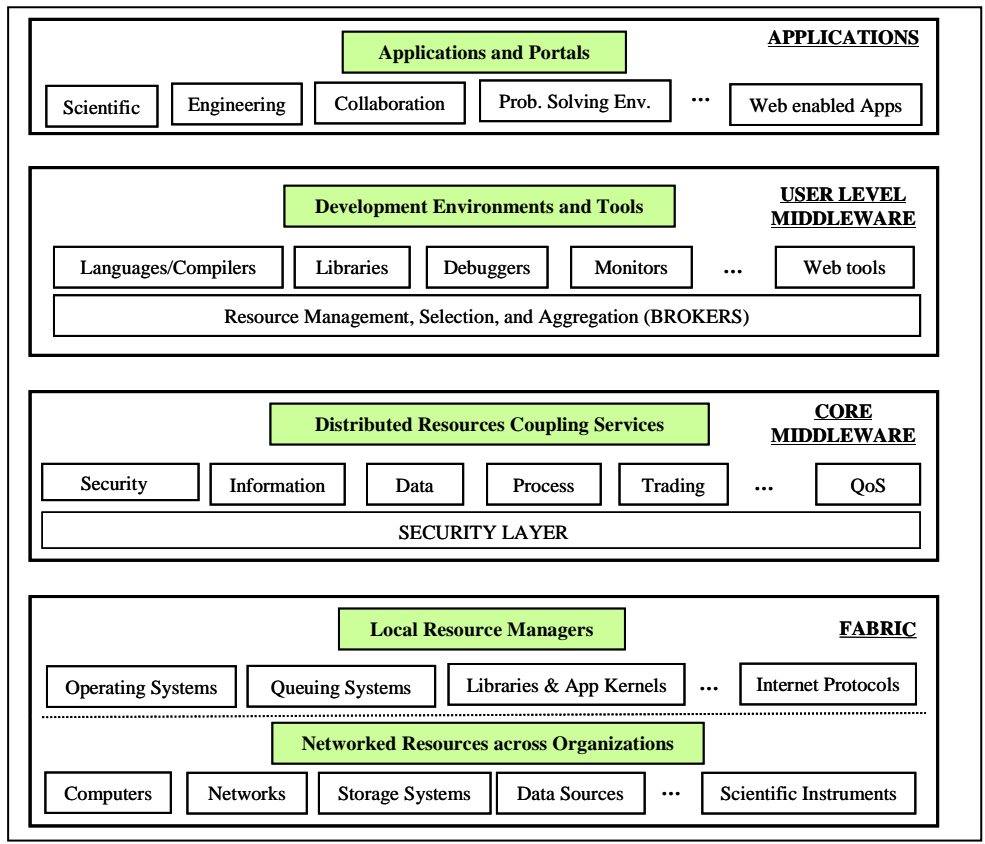

Figure 2: A Layered Grid Architecture and components.

*Core Grid middleware* offers services such as remote process management, co-allocation of resources, storage access, information registration and discovery, security, and aspects of Quality of Service (QoS) such as resource reservation and trading. These services abstract the complexity and heterogeneity of the fabric level by providing a consistent method for accessing distributed resources.

*User-level Grid middleware* utilizes the interfaces provided by the low-level middleware to provide higher level abstractions and services. These include application development environments, programming tools and resource brokers for managing resources and scheduling application tasks for execution on global resources.

*Grid applications and portals* are typically developed using Grid-enabled languages and utilities such as HPC++ or MPI. An example application, such as parameter simulation or a grand-challenge problem, would require computational power, access to remote data sets, and may need to interact with scientific instruments. Grid portals offer Web-enabled application services, where users can submit and collect results for their jobs on remote resources through the Web.

The middleware surveyed in this chapter extend across one or more of the levels above the Grid fabric layer of this generic stack. A short description for each of them is provided in Table 1.

| Name | Description | Remarks | Website |
|---|---|---|---|
| UNICORE | Vertically integrated Java based Grid computing environment that provides a seamless and secure access to distributed resources. | Project funded by the German Ministry for Education and Research with co-operation between ZAM, Deutscher, etc | http://www.unicore.org |
| Globus | Open source software toolkit that facilitates construction of computational grids and grid based applications, across corporate, institutional and geographic boundaries without | R&D project conducted by the "Globus Alliance" which includes Argonne National Laboratory, Information Sciences Institute and others. | http://www.globus.org |

| | sacrificing local autonomy. | | |
|---|---|---|---|
| Legion | Vertically integrated Object-based metasystem that helps in combining a large numbers of independently administered heterogeneous hosts, storage systems, databases legacy codes and user objects distributed over wide-area-networks into a single, object-based *metacomputer* that accommodates high degrees of flexibility and site autonomy. | A R&D project at the University of Virginia, USA. The software developed by this project is commercialized through a new company called Avaki. | http://legion.virginia.edu |
| Gridbus | Open source software toolkit that extensively leverages related software technologies and provides an abstraction layer to hide idiosyncrasies of heterogeneous resources and low-level middleware technologies from application developers. It focuses on realization of utility computing and market-oriented computing models scaling from clusters to grids and to peer-to-peer computing systems. | A research and innovation project led by the University of Melbourne GRIDS Lab with support from the Australian Research Council. | http://www.gridbus.org/ |

**Table 1: Grid middleware systems.**

## 3. UNICORE

UNICORE [13] is a vertically integrated Grid computing environment that facilitates the following:
- A seamless, secure and intuitive access to resources in a distributed environment – for end users.
- Solid authentication mechanisms integrated into their administration procedures, reduced training effort and support requirements – for Grid sites.
- Easy relocation of computer jobs to different platforms – for both end users and Grid sites.

UNICORE follows a three-tier architecture, which is shown in Figure 3 (drawn with ideas from [14]). It consists of a client that runs on a Java enabled user workstation or a PC, a gateway, and multiple instances of Network Job Supervisors (NJS) that execute on dedicated securely configured servers and multiple instances of Target System Interfaces (TSI) executing on different nodes provide interfaces to underlying local resource management systems such as operating systems and the batch subsystems. From an end user's point of view, UNICORE is a client-server system based on a three-tier model:

- User tier: The user is running the UNICORE Client on a local workstation or PC.
- Server tier: On the top level, each participating computer center defines one or several UNICORE Grid sites (*Usites*) that Clients can connect to.
- Target System tier: A *Usite* offers access to computing or data resources. They are organized as one or several virtual sites (*Vsites*) which can represent the execution and/or storage systems at the computer centers.

The UNICORE Client interface consists of two components: JPA (Job Preparation Agent) and JMC (Job Monitor Component). Jobs are constructed using JPA and the status and results of the jobs can be obtained through the JMC. The jobs or status requests and the results are formulated in an abstract form using the *Abstract Job Object (AJO)* Java classes. The client connects to a UNICORE *Usite* gateway and submits the jobs through AJOs.

The UNICORE Gateway is the single entry point for all UNICORE connections into a *Usite*. It provides an Internet address and a port that users can use to connect to the gateway using SSL.

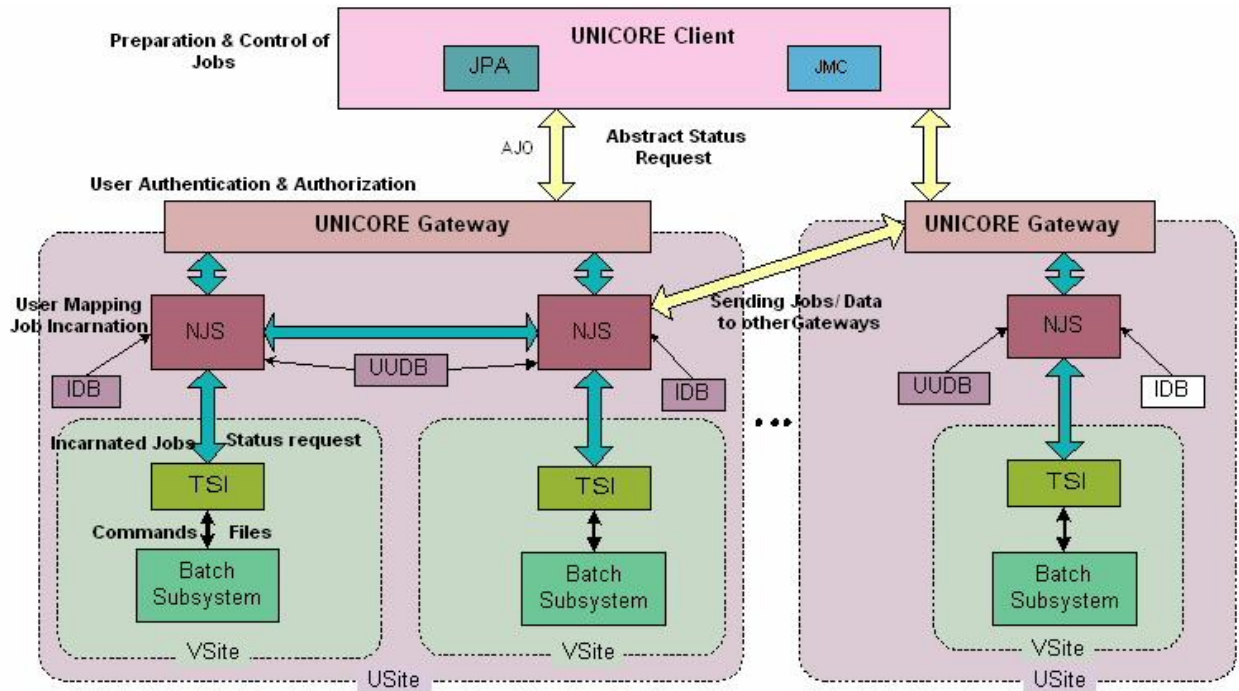

**Figure 3: The UNICORE Architecture.**

A UNICORE *Vsite* is made up of two components: NJS (Network Job Supervisor) and TSI (Target System Interface). The NJS Server manages all submitted UNICORE jobs and performs user authorization by looking for a mapping of the user certificate to a valid login in the UUDB (UNICORE User Data Base). NJS also deals with the incarnation of jobs from the AJO definition into the appropriate concrete command sequences for a given target execution system, based on specifications in the Incarnation Data Base (IDB). UNICORE TSI accepts incarnated job components from the NJS, and passes them to the local batch systems for execution.

UNICORE's features and functions can be summarized as follows:

1. **User driven job creation and submission**: A graphical interface assists the user in creating complex and interdependent jobs that can be executed on any UNICORE site without job definition changes.
2. **Job management**: The Job management system provides user with full control over jobs and data.
3. **Data management**: During the creation of a job, the user can specify which data sets have to be imported into or exported from the *USpace* (set of all files that are available to a UNICORE job), and also which datasets have to be transferred to a different *USpace*. UNICORE performs all data movement at run time, without user intervention.
4. **Application support**: Since scientists and engineers use specific scientific application, the user interface is built in pluggable manner in order to extend it with plugins that allows to prepare specific application input.
5. **Flow control**: A user job can be described as a set of one or more directed acyclic graphs.
6. **Single sign-on**: UNICORE provides a single sign-on through X.509V3 certificates.
7. **Support for legacy jobs**: UNICORE supports traditional batch processing by allowing users to include their old job scripts as part of a UNICORE job.
8. **Resource management**: Users select the target system and specify the required resources. The UNICORE client verifies the correctness of jobs and alerts users to correct errors immediately.

The major Grid tools and application projects making use of UNICORE as their low-level middleware include: EuroGrid [18] and their applications—BioGrid [19], MeteoGrid, and CAEGrid, Grid Interoperability Project (GRIP)[20], OpenMolGrid [19], and Japanese NAREGI (National Research Grid Initiative) [22].

**Overview of Job Creation, Submission and Execution in UNICORE Middleware**

The UNICORE Client assists in creating, manipulating and managing complex, interdependent multi-system jobs, multi-site jobs, synchronization of jobs and movement of data between systems, sites and storage spaces. The client creates an Abstract Job Object (AJO) represented as a serialized Java Object or in XML format. The UNICORE Server (NJS) performs

- Incarnation of the AJO into target system specific actions
- Synchronization of actions (work flow)
- Transfers of jobs and data between User Workstation, Target Systems and other sites
- Monitoring of status

The two main areas of UNICORE are: 1) seamless specification of some work to be done at a remote site and 2) transmission of the specification, results and related data. The seamless specification in UNICORE is dealt with by a collection of Java classes known loosely as the AJO (Abstract Job Object) and the transmission is defined in the UNICORE Protocol Layer (UPL). The UPL is designed as a protocol that transmits data regardless of its form. The classes concerned with the UPL are included in the org.unicore.package and the org.unicore.upl package, with some auxiliary functions from the org.unicore.utility package.

The main packages for the AJO are org.unicore.ajo, org.unicore.outcome, org.unicore.resources and org.unicore.idiomatic. The AJO is how a UNICORE client application such as a Job Preparation Agent (JPA) or Job Monitor Controller (JMC), can specify the work that a UNICORE User wants to do, to a UNICORE server (NJS). One aim of the AJO is to allow seamless specification of jobs that can be retargeted to different computer sites just by changing the address, the specification of the work remains the same – regardless of differing site policies, different executables and options, different authorisation policies etc. The mapping of the seamless specification to the actual site values, known as *incarnation* in UNICORE, is done at run time by the NJS at the target site.

The atom of execution in UNICORE is an *AbstractJob*. NJSs execute AbstractJobs for users. An Abstract Job contains one or more sub-tasks. The sub-tasks are Abstract Actions and define simple actions, such as fetch files, compile, execute. Abstract Jobs are also Abstract Actions and so can be contained within a parent Abstract Job.

When an NJS starts executing an Abstract Job, it creates a directory on a file system for the Abstract Job. This directory is known as the Abstract Job's *Uspace*. All files used by the child Abstract Actions are assumed to be in the *Uspace*. Most Abstract Actions do not have direct access to a site's file system (**Xspace**). Any file used by Abstract Actions has to be imported to the *Uspace* before they are used and any files that have to be saved have to be explicitly saved from the *Uspace*. The available ways to save a file are to export it, return it with the AJO's results or to spool it to a semi-permanent holding area. The *Uspace* is deleted when the AJO finishes execution.

The child Abstract Actions of an Abstract Job are kept in a Directed Acyclic Graph (DAG) which defines the order of execution. Successor Abstract Actions start execution when all their predecessor Abstract Actions have finished execution successfully. If an Abstract Action fails in execution, then none of its successor Abstract Actions are executed.

The results of executing Abstract Actions are known as Outcomes. Outcomes consist of the stdout and stderr produced by an executable, logging and status information from the NJS and/or representations of the results returned by the Abstract Action. Each type of Abstract Action is matched by a type of Outcome.

The core classes for the definition of work are the subclasses of org.unicore.ajo.AbstractTask. These define work that will be performed by the target system, outside of the NJS, and so define tasks such as execute, link, compile and various file manipulations. These tasks can request resources such as the number of processors, amount of memory, time and also any external software packages they require. A client can ask the site for a description of the supported resources and so only request resources that are available.

The other major grouping of Abstract Actions is based on those that can manipulate other Abstract Actions e.g. return current status, return results, kill etc

The UNICORE Object hierarchy is shown below:
- AbstractAction: Parent class of all UNICORE actions.
- ActionGroup: Container for UNICORE actions.
- AbstractJob: ActionGroup which can run remotely.
- RepeatGroup: Actions in a loop.
- AbstractTask: A computational action, e.g. copy file.
- AbstractService: A service action, e.g. kill job.
- ConditionalAction: If-then-else for actions.

ActionGroup contains a DAG of AbstractActions which shows dependencies between actions (nodes) that define control flow. Actions in the DAG can be any subtype of AbstractAction. An action starts when all it predecessors are "DONE". The following subclasses of "DONE" are used for control.
- SUCCESSFUL: The AbstractAction completed without error.
- NOT_SUCCESSFUL: The AbstractAction failed.
- NEVER_RUN: A predecessor of the AbstractAction failed.

- NEVER_TAKEN: The AbstractAction is on the not taken branch of a conditional action.

The work flow constructs in UNICORE allow:
- Automating complex multi-site, multi-system chains of Jobs
- Run computational experiments like parameter studies
- Use all features of UNICORE, like security and seamlessness

## 4. Globus

The Globus project provides open source software toolkit [13] that can be used to build computational grids and grid based applications. It allows sharing of computing power, databases, and other tools securely online across corporate, institutional and geographic boundaries without sacrificing local autonomy. The core services, interfaces and protocols in the Globus toolkit allow users to access remote resources seamlessly while simultaneously preserving local control over who can use resources and when. The Globus architecture, shown in Figure 4, has three main groups of services accessible through a security layer. These groups are Resource Management, Data management and Information Services.

The local services layer contains the operating system services, network services like TCP/IP, cluster scheduling services provided by Load Leveler, job-submission, query of queues, and so on. The higher layers of the Globus model enable the integration of multiple or heterogeneous clusters. The core services layer contains the Globus toolkit building blocks for security, job submission, data management, and resource information management. The high-level services and tools layer contains tools that integrate the lower level services or implement missing functionality.

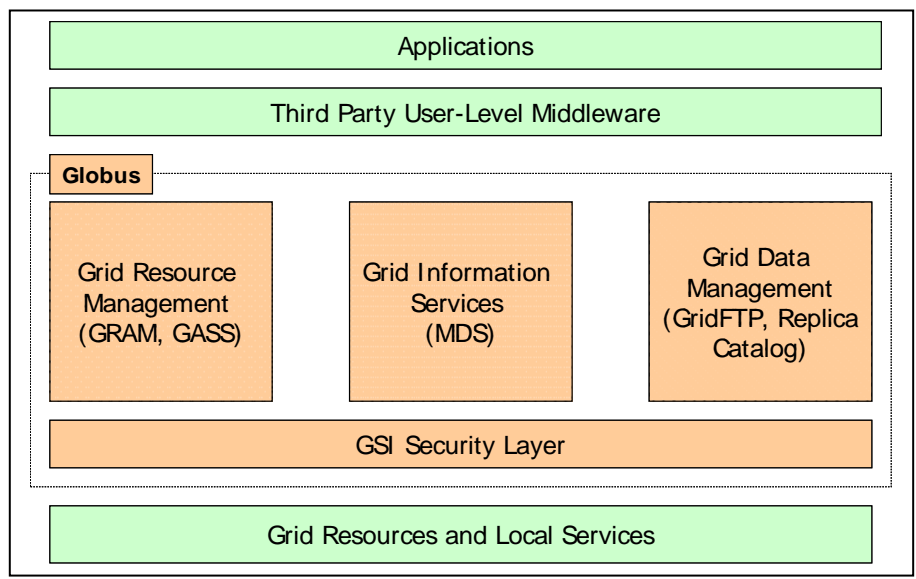

**Figure 4: The Globus Architecture.**

### GSI Security Layer:

The Grid Security Infrastructure (GSI) [23] provides methods for authentication of Grid users and secure communication. It is based on SSL (Secure Sockets Layer), PKI (Public Key Infrastructure) and X.509 Certificate Architecture. The GSI provides services, protocols and libraries to achieve the following aims for Grid security:
- Single sign-on for using Grid services through user certificates
- Resource authentication through host certificates
- Data encryption
- Authorization
- Delegation of authority and trust through proxies and certificate chain of trust for Certificate Authorities (CAs)

Users gain access to resources by having their Grid certificate subjects mapped to an account on the remote machine by its system administrators. This also requires that the CA that signed the user certificate be trusted by the remote system. Access permissions have to be enforced in the traditional UNIX manner through restrictions on the remote user account.

CAs are also a part of realising the notion of Virtual Organizations (VOs) [54]. VOs are vertical collaborations and a user who has a certificate signed by the CA of the VO gains access to the resources authenticated by the same CA. VOs can cooperate between themselves by recognizing each others CAs so that users can access resources between collaborations. These mechanisms are used in many Grid testbeds. Depending on the structure of the testbed and the tools used, the users may gain access automatically to the resources or may have to contact the system administrators individually to ensure access.

Most services require mutual authentication before carrying out their functions. This guarantees non-repudiability and data security on both sides. However, the current state of GSI tools makes it more likely that some users may share the usage of a single certificate to gain access to higher number of resources or that they may be mapped to the same account on the remote machine. This may raises serious questions on the authenticated users and the confidentiality of user data on the remote machine. Production testbeds have policies in place to restrict this behaviour but there is still some way to go before these are restricted at the middleware level.

**Resource Management:**

The resource management package enables resource allocation through job submission, staging of executable files, job monitoring and result gathering. The components of Globus within this package are:

**Globus Resource Allocation Manager (GRAM):** GRAM [24] provides remote execution capability and reports status for the course of the execution. A client requests a job submission to the gatekeeper daemon on the remote host. The gatekeeper daemon checks if the client is authorized (i.e., the client certificate is in order and there is a mapping of the certificate subject to any account on the system). Once authentication is over, the gatekeeper starts a job manager that initiates and monitors the job execution. Job managers are created depending on the local scheduler on that system. GRAM interfaces to various local schedulers such as Portable Batch System (PBS), Load Sharing Facility (LSF) and LoadLeveler.

The job details are specified through the Globus Resource Specification Language (RSL), which is a part of GRAM. RSL provides syntax consisting of attribute-value pairs for describing resources required for a job including the minimum memory and the number of CPUs.

**Globus Access to Secondary Storage (GASS):** GASS [27] is a file-access mechanism that allows applications to pre-fetch and open remote files and write them back. GASS is used for staging-in input files and executables for a job and for retrieving output once it is done. It is also used to access the standard output and error streams of the job. GASS uses secure HTTP based streams to channel the data and has GSI-enabled functions to enforce access permissions for both data and storage.

**Information Services:**

The information services package provides static and dynamic properties of the nodes that are connected to the Grid. The Globus component within this package is called *Monitoring and Discovery Service (MDS)* [25].

MDS provides support for publishing and querying of resource information. Within MDS, schema define classes that represent various properties of the system. MDS has a three-tier structure at the bottom of which are Information Providers (IPs) that gather data about resource properties and status and translate them into the format defined by the object classes. The Grid Resource Information Service (GRIS) forms the second tier and is a daemon that runs on a single resource. GRIS responds to queries about the resource properties and updates its cache at intervals defined by the time-to-live by querying the relevant IPs. At the topmost level, the GIIS (Grid Information Index Service) indexes the resource information provided by other GRISs and GIISs that are registered with it.

The GRIS and GIIS run on Lightweight Directory Access Protocol (LDAP) backend in which the information is represented as a hierarchy of entries, each entry consisting of 0 or more attribute-value pairs. The standard set of IPs provide data on CPU type, system architecture, number of processors and memory available among others.

**Data Management:**

The data management package provides utilities and libraries for transmitting, storing and managing massive data sets that are part and parcel of many scientific computing applications [26]. The elements of this package are:

*GridFTP:* It is an extension of the standard FTP protocol that provides secure, efficient and reliable data movements in grid environments. In addition to standard FTP functions, GridFTP provides GSI support for authenticated data transfer, third-party transfer invocation and striped, parallel and partial data transfer support.

*Replica Location and Management:* This component supports multiple locations for the same file throughout the grid. Using the replica management functions, a file can be registered with the Replica Location Service (RLS) and its replicas can be created and deleted. Within RLS, a file is identified by its Logical File Name (LFN) and is registered within a logical collection. The record for a file points to its physical locations. This information is available from the RLS upon querying.

The major Grid tools and application projects making use of Globus as their low-level middleware include: AppLeS [28], Ninf [30], Nimrod-G [29], NASA IPG[36], Condor-G [31], Gridbus Broker [32], UK eScience Project [33], GriPhyN [35], and EU Data Grid [34].

## 5. Legion

Legion [16] is a middleware system that combines very large numbers of independently administered heterogeneous hosts, storage systems, databases legacy codes and user objects distributed over wide-area-networks into a single coherent computing platform. Legion provides the means to group these scattered components together into a single, object-based metacomputer that accommodates high degrees of flexibility and site autonomy.

Figure 5 shows the architecture for Legion middleware. It is structured as a system of distributed "objects" – active processes that communicate using a uniform remote method invocation service. All hardware and software resources in a grid system will be represented by *Legion objects*. Legion's fundamental object models are described using an interface description language (IDL), and are compiled and linked to implementations in a given language. This approach enables component interoperability between multiple programming languages and heterogeneous execution platforms. Since all elements in the system are objects, they can communicate with one another regardless of location, heterogeneity, or implementation details thereby addressing problems of encapsulation and interoperability.

A "class object" is used to define and manage its corresponding Legion object. Class objects are given system-level responsibility; it controls the creation of new instances, scheduling execution, activating and deactivating instances, and provides information about their current location to client objects that wish to communicate with the instances. In other words, classes are act as *managers* and *policy makers* of the system. Metaclasses are used to describe the classes' instances.

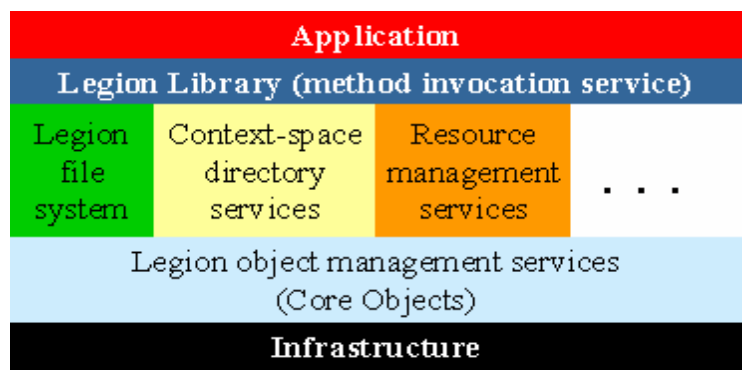

**Figure 5: Legion Architecture.**

Legion defines a set of core object types that support basic system services, such as naming and binding, object creation, activation, deactivation, and deletion. These objects provide the mechanisms that help classes to implement policies appropriate for their instances. Legion also allows users to define and build their own class objects. Some core objects are:

- *Host objects*: represent processors in Legion.
- *Vault objects*: represent persistent storage.
- *Context objects*: map context names to LOIDs (Legion Object Identifiers).
- *Binding agents*: LOIDs to LOAs (Legion Object Address).
- *Implementation object*: maintains as an executable file that a host object can execute when it receives a request to activate or create an object.

Host objects provide a uniform interface to object (task) creation, and vault objects provide a uniform storage allocation interface, even though there may be many different implementations of each of these. Also, these objects naturally act as resource guardians and policy makers.

A three-level naming system is used in Legion. Human-readable strings, called "context names", which allow users to take advantage of a wide range of possible resources, are at the highest level. Context objects map the context names to LOIDs (Legion Object Identifiers) that forms the next level. LOIDs are location independent and hence are insufficient for communication; therefore, LOIDs are transformed into LOAs (Legion Object Addresses) for communication. A LOA is a physical address (or set of addresses in the case of a replicated object) that contains sufficient information to allow other objects to communicate with the object.

The major Grid tools and testbeds that made or are making use of Legion as their low-level middleware include: NPACI Testbed [42], Nimrod-L [41], and NCBioGrid [40]. Additionally, it has been used in the study of axially symmetric steady flow [39] and protein folding [38] applications.

## 6. Gridbus

The Gridbus Project [17] is an open-source, multi-institutional project led by the GRIDS Lab at the University of Melbourne. It is engaged in the design and development of service-oriented cluster and grid middleware technologies to support eScience and eBusiness applications. It extensively leverages related software technologies and provides an abstraction layer to hide idiosyncrasies of heterogeneous resources and low-level middleware technologies from application developers. In addition, it extensively focuses on realization of utility computing model scaling from clusters to grids and to peer-to-peer computing systems. It uses economic models [43] in efficient management of shared resources and promotes commoditization of their services. Thus, it enhances the tradability of grid services and manages efficiently the supply and demand for resources.

Gridbus supports commoditization of grid services at various levels:
- Raw resource level (e.g., selling CPU cycles and storage resources)
- Application level (e.g., molecular docking operations for drug design application [7] )
- Aggregated services (e.g., brokering and reselling of services across multiple domains)

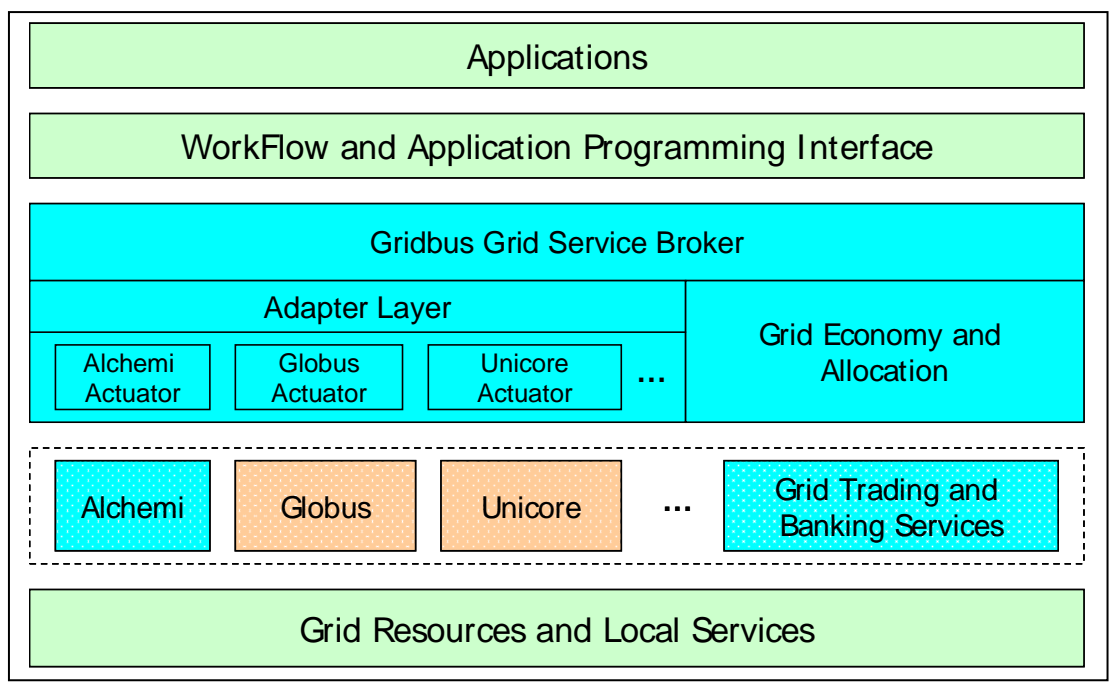

**Figure 6: The Gridbus Architecture.**

The idea of a computational economy helps in creating a service-oriented computing architecture where service providers offer paid services associated with a particular application and users, based on their requirements, would optimize by selecting the services they require and can afford within their budget. Gridbus emphasizes the end-to-end quality of services driven by computational economy at various levels - clusters, peer-to-peer (P2P) networks, and the Grid - for the management of distributed computational, data, and application services.

Figure 6 shows a layered architecture depicting the Gridbus components in conjunction with other middleware technologies—Globus and Unicore—that have been discussed before and Alchemi which is discussed in detail in the next chapter, although we briefly present it here for completeness.

Gridbus provides software technologies that spread across the following categories:
- Enterprise Grid Infrastructure (Alchemi)
- Cluster Economy and Resource Allocation (Libra)
- Grid Economy and Virtual Enterprise (Grid Market Directory, Compute Power Market)

- Grid Trading and Accounting Services (GridBank)
- Grid Resource Brokering and Scheduling (Gridbus Broker)
- Grid Workflow Management (Gridbus Workflow Engine)
- Grid Application Programming Interface (Visual Parametric Modeller)
- Grid Portals (GMonitor, Gridscape)
- Grid Simulation (GridSim)

*Alchemi*: Though scientific computing facilities have been heavy users of Unix-class OSes, the vast majority of computing infrastructure within enterprises is still based on Microsoft Windows. Alchemi was developed to address the need within enterprises for a desktop grid solution that utilizes the unused computational capacity represented by the vast number of PCs and workstation running Windows within an organization. Alchemi is implemented on top of the Microsoft .NET Framework and provides the runtime machinery for constructing and managing desktop grids. It also provides an object-oriented programming model along with web service interfaces that enable its services to be accessed from any programming environment that supports SOAP-XML abstraction.

*Libra*: Libra is a cluster scheduling system that guarantees a certain share of the system resources to a user job such that the job is completed by the deadline specified by the user provided he has the requisite budget for it. Jobs whose output is required immediately require a higher budget than those with a more relaxed deadline. Thus, Libra delivers utility value to the cluster users and increases their satisfaction by creating realistic expectations for the job turnaround times.

*Market Mechanisms for Computational Economy:* Grid Market Directory (GMD) is a registry service where service providers can register themselves and publish the services they're providing and consumers can query to obtain the service that meets their requirements. Some of the attributes of a service are its access point, input mechanism and the cost involved in using it.

Compute Power Market (CPM) is a market-based resource management and scheduling system developed over the JXTA platform. It enables trading of idle computational power over P2P networks. The CPM components that represent markets, consumers and providers are Market Server, Market Resource Agent, and Market Resource Broker (MRB). It supports various economic models for resource trading and matching service consumers and providers and allows plugging in of different scheduling mechanisms.

*Accounting and Trading Services:* GridBank is a Grid-wide accounting and micro-payment service that provides a secure infrastructure for Grid Service Consumers (GSCs) to pay Grid Service Providers (GSPs) for the usage of their services. The consumer is charged on the basis of resource usage records maintained by the provider and service charges that have been agreed upon by both parties in the beginning. GridBank can also be used as an authentication and authorization mechanism thereby ensuring access to the resources to only those consumers with the requisite credit in their accounts.

*Resource Broker:* The Gridbus Resource Broker provides an abstraction to the complexity of Grids by ensuring transparent access to computational and data resources for executing a job on a grid. It uses user requirements to create a set of jobs, discover resources, schedule, execute and monitor the jobs and retrieve their output once they are finished. The broker supports a declarative and dynamic parametric programming model for creating grid applications.

The Gridbus broker has the capability to locate and retrieve the required data from multiple data sources and to redirect the output to storage where it can be retrieved by processes downstream. It has the ability to select the best data repositories from multiple sites based on availability of files and quality of data transfer.

*Web Portals:* G-monitor is a web-portal for monitoring and steering computations on global grids. G-monitor interfaces with resource brokers such as Gridbus broker and Nimrod-G and uses their services to initiate and monitor application execution. It provides the user with up-to-date information about the progress of the execution at the individual job level and at the overall experiment level. At the end of the execution, the user can collect the output files through G-monitor.

To manage and monitor Grid testbeds, the Gridbus Project has created a testbed portal and generation tool called Gridscape. Gridscape generates interactive and dynamic portals that enable users to view the status of the resources within the testbed and easily add new resources when required. It is also possible to customize the portal to reflect the unique identity of the organization managing the testbed.

*Simulation and Modeling:* The GridSim toolkit provides facilities for the modeling and simulation of resources and network connectivity with different capabilities, configurations and domains. It supports primitives for application composition, information services for resource discovery and interfaces for assigning application tasks to resources and managing their execution. It also provides a visual modeler interface for creating users and resources. These features can be used to simulate parallel and distributed scheduling systems such as resource brokers or Grid schedulers for evaluating performance of scheduling algorithms or heuristics.

The major Grid tools and application projects making use of Gridbus components within their middleware include: ePhysics Portal [52], Australian Virtual Observatory[51], Belle Analysis Data Grid[50], Global Data Intensive Grid Collaboration [49], NeuroGrid [48], Natural Language Engineering on the Grid [53], HydroGrid [46], and Amsterdam Private Grid [47].

## 7. Implementation of UNICORE Adaptor for Gridbus Broker

As a complete integrated system, Legion provides its own grid broker that facilitates management, selection and aggregation of resources and scheduling of applications for execution on global resources whereas Globus and UNICORE do not provide their own. However, Globus has a number of third part grid broker implementations such as Nimrod-G, Gridbus Broker, and Condor-G, but UNICORE does not have such an add-on service. The absence of brokerage services in UNICORE motivates the need for porting a Grid resource broker to UNICORE. An extension of Gridbus broker and the implementation of adaptor for UNICORE middleware is described in this section.

The scheduler in Gridbus has a platform-independent view of the nodes and is concerned only with their performance. Therefore, it is possible for the broker to operate across different middleware. The Gridbus broker has already capable of operating on resources which are grid-enabled using Globus and Alchemi middleware. Similarly, the Gridbus Broker can be designed to work with UNICORE middleware by extending the ComputeServer, JobWrapper, JobMonitor, and JobOutput classes in Gridbus. The main components in the implementation of Gridbus broker on UNICORE is shown in the UML Diagram in Figure 8.

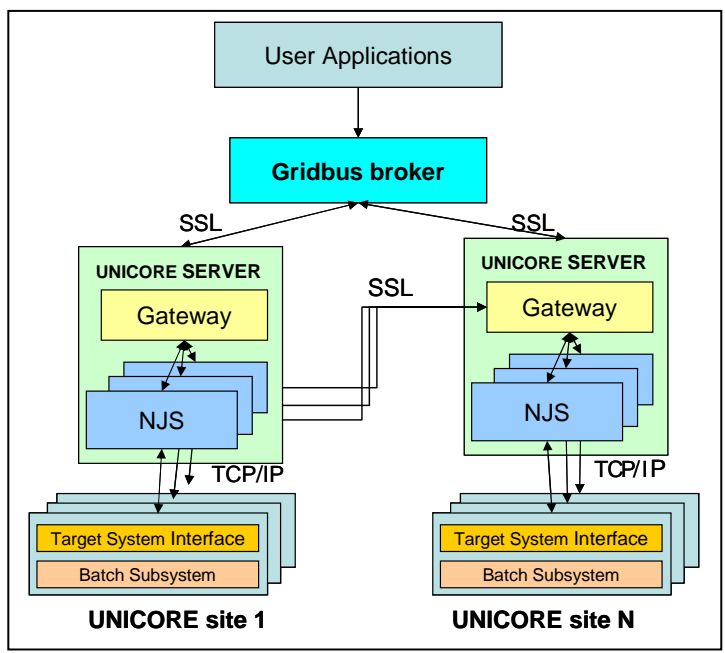

**Figure 7: Schematic for interfacing Gridbus broker on UNICORE middleware.**

### UnicoreComputeServer

The ComputeServer object in Gridbus broker is responsible for describing nodes in a system. By extending the ComputeServer class for UNICORE, the attributes of nodes that are grid-enabled by UNICORE can be described in UnicoreComputeServer.

In order to establish access to resources, first we need to create an identity that is used for AJO execution.
```
Identity identity = new Identity( new File(<keystore>), <password>);
```

The created identity is then used to set up a SSL connection to the gateway and to establish a link to the *Vsite*.
```
Reference reference = new Reference.SSL(<gateway_address>, identity);
VsiteTh vsite = new VsiteTh(reference, <Vsite name>);
```

Since the information is extracted from the command line directly, no checking is required by the gateway. Dynamic applications can only accept some gateway addresses and then present the list of *Vsites* and ports to the user. Resources could also be fetched from the *Vsites* by using the *VsiteManager* that provides a global view of all the known *Vsites*.

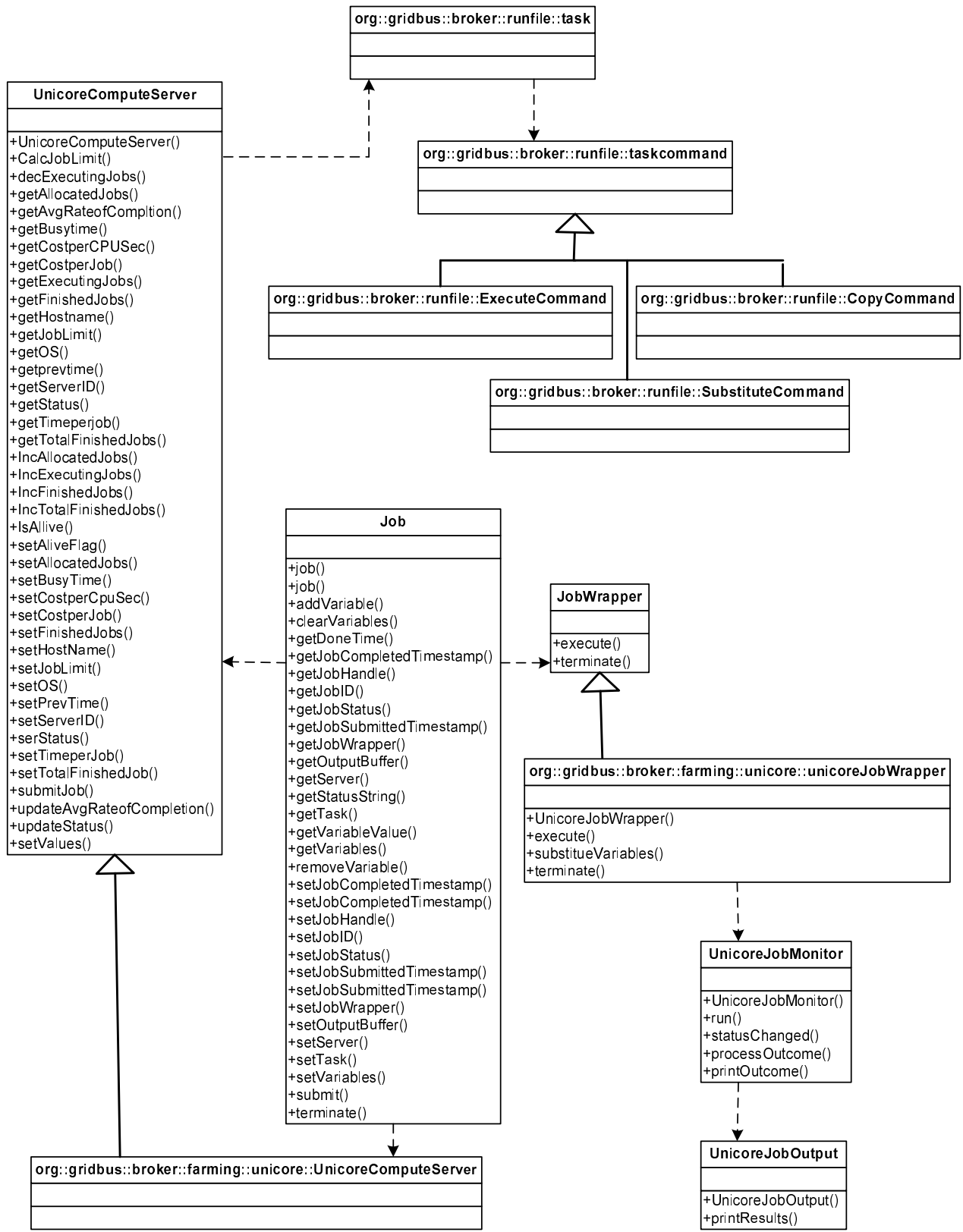

**Figure 8: Main Components of Gridbus-UNICORE interface.**

## UnicoreJobWrapper

The main function of this object is to convert the job specified by the Gridbus broker through an XML-based Parameter Sweep Specification (xPSS) document to a format which is understandable by the ComputeServer running UNICORE middleware. In xPSS document, there are mainly three types of commands namely, SUBSTITUTE, COPY and EXECUTE. For the COPY command, we need to copy the files required for the execution of AJO from the client machine to the remote machine and copy the result files back to the local machine.

In the first case, the required files are first put into the IncarnateFiles which creates files in the *Uspace* from data contained in the instance. This is provided as an alternative way of transferring files to *Uspaces*. In this case the bytes of the files are placed in the AJO and therefore, this method is only practical for small files. Large files should be streamed. At last, files are put into a portfolio which is transferred along with the AJO to be executed.

**Method 1:**

```
MakePortfolio mp1= new MakePortfolio ();
File file=new File(<file_name>);
FileInputStream fin=new FileInputStream(file);
byte[] file_contents= new byte[(int) file.length()];
fin.read(file_contents);
IncarnateFiles inf1 = new IncarnateFiles("Required Files");
inf1.addFile(<file_name>,file_contents);
mp1.addFile(<file_name>);
```

**Method 2:**

```
Portfolio portfolio = new Portfolio();
PortfolioTh files = new PortfolioTh(portfolio, <java.io.File[] Files>);
```

In the second case, the result files in the remote machine are added to the job Outcome which is transferred to the local machine. This is done because all the files in the *Uspace* are deleted at the end of AJO execution.

Step 1: let the NJS know about the files (make a Portfolio)
```
MakePortfolio mp2 = new MakePortfolio()
mp2.addFile(<result file name in Upsace>);
```

Step 2: save files
```
CopyPortfolioToOutcome cpto = new CopyPortfolioToOutcome ();
cpto.setTarget(mp2.getPortfolio().getId());
```

For the EXECUTE command, the executable in the plan file is converted into a script and an `ExecuteScriptTask` is created which is included in the AJO.

```
String the_script = "date\n";
the_script += "hostname\n";
inf.addFile("script",the_script.getBytes());

MakePortfolio mp = new MakePortfolio("AJO Example");
mp.addFile("script");

ExecuteScriptTask est = new ExecuteScriptTask("AJO Example Task");
est.setScriptType(ScriptType.CSH);
est.setExecutable(mp.getPortfolio());
```

Finally, the AJO is created by adding the components and their dependencies and is submitted to the remote machine using `job_manager.consignSynchronous` method in com.fujitsu.arcon.servlet.JobManager class within the Arcon client library.

```
AbstractJob ajo = new AbstractJob("AJO Example");

ajo.addDependency(inf,mp);
ajo.addDependency(inf1,mp1);
ajo.addDependency(mp,est);
ajo.addDependency(est, mp2);
```

```
        ajo.addDependency(mp2,cpto);
        outcome = JobManager.consignSynchronous(ajo,vsite);
```

The UnicoreJobWrapper also starts the UnicoreJobMonitor thread which is responsible for monitoring the status of the job execution.

### UnicoreJobMonitor

The UnicoreMonitor is used to monitor the execution of the job and display the detailed status information during the execution. The status information includes those related to all AbstractActions and AbstractTasks involved such as MakePortfolio and ExecuteScriptTask. It is also responsible for terminating the job in case of job failure or job completion.

### UnicoreJobOutput

As the name suggests, UnicoreJobOutput is used to get back the results of the job. It displays the contents of the standard output and standard error files from the computeServer on the local machine. If other result files are produced as a result of the job execution, UnicoreJobOutput renames these files to "filename.jobid" format and saves them in the local directory. The Outcome is fetched from the client side using the `Job_manager.getOutcome` function of the JobManager. Thus, all the result files are written to local storage if the AJO is successfully completed.

```
        Collection c =  outcome.getFilesMapping().get(cpto.getId());
        Iterator i = c.iterator();
        while (i.hasNext())
        {
                File f = (File)i.next();
                System.out.println("\n\nRESULT FILE:
                   "+f.getCanonicalPath()+"\nCONTENTS\n");
                BufferedReader reader = new BufferedReader(new FileReader(f));
                String line;
                while((line = reader.readLine()) != null)
                {System.out.println(line);}
        }
```

## 7. Comparison of Middleware Systems

Figure 9 compares the surveyed middleware on the basis of the services provided across the grid architecture stack. UNICORE and Legion are vertically integrated and tightly coupled. They provide both server and client components. Globus follows a "bag-of-services" approach and provides rich set of basic tools that can be used selectively to construct grid systems. For example, although Gridbus has its own independent low-level middleware, its broker has been designed to operate with Globus. The Gridbus component set, though spread across the stack, are not integrated as tightly as UNICORE and Legion and can be used completely independent of each other.

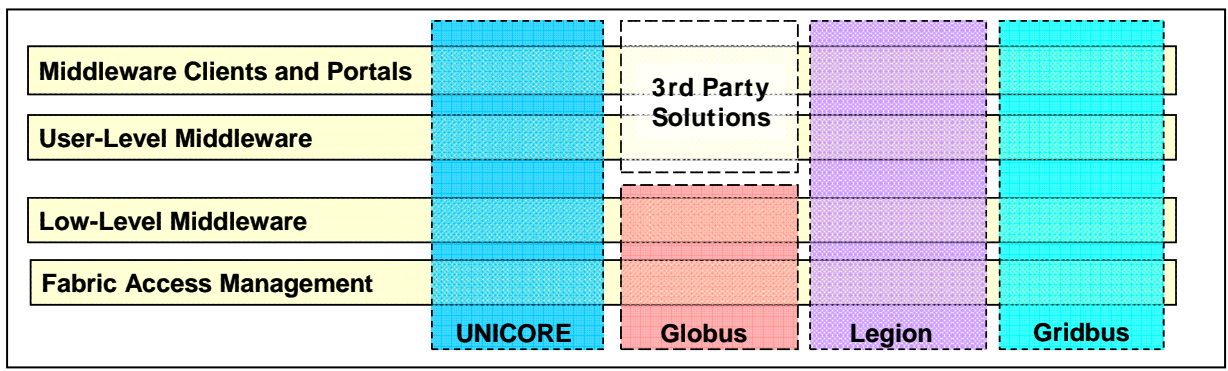

**Figure 9: Comparison of UNICORE, Globus, Legion and Gridbus**

A comparison of the various middleware systems based on their architecture, implementation model and category is given in Table 2. Another comparison between Globus and Legion can be found in [37].

| Middleware Property | **UNICORE** | **Globus** | **Legion** | **Gridbus** |
|---|---|---|---|---|
| Focus | High level Programming models | Low level services | High level Programming models | Abstractions and market models |
| Category | Mainly uniform job submission and monitoring | Generic computational | Generic computational | Generic computational |
| Architecture | Vertical multi tiered system | Layered and modular toolkit | Vertically integrated system | Layered component and utility model |
| Implementation Model | Abstract Job Object | Hourglass model at system level | Object-oriented metasystem | Hourglass model at user level |
| Implementation Technologies | Java | C and Java | C++ | C, Java, C# and Perl |
| Runtime Platform | Unix | Unix | Unix | Unix and Windows with .NET (Alchemi) |
| Programming Environment | Workflow environment | Replacement libraries for Unix & C libraries. Special MPI library (MPICH –G), CoG (Commodity Grid) kits in Java, Python, CORBA, Matlab, Java Server Pages, Perl and Web Services | Legion Application Programming Interfaces (API). Command line utilities | Broker Java API XML-based parameter-sweep language Grid Thread model via Alchemi. |
| Distribution Model | Open source | Open source | Not open source. Commercial version available | Open source |
| Some Users and Applications | EuroGrid [18], Grid Interoperability Project (GRIP) [20], OpenMolGrid [19], and Japanese NAREGI [22]. | AppLeS [28], Ninf [30], Nimrod-G [29], NASA IPG [36], Condor-G [31], Gridbus Broker [32], UK eScience Project [33], GriPhyN [35], and EU Data Grid [34]. | NPACI Testbed [42], Nimrod-L [41], and NCBioGrid [40]. Additionally, it has been used in the study of axially symmetric steady flow [39] and protein folding [38] applications. | ePhysics Portal [52], Belle Analysis Data Grid[50], NeuroGrid [48], Natural Language Engineering [53], HydroGrid [46], and Amsterdam Private Grid [47]. |

**Table 2: Comparison of Grid Middleware Systems.**

In addition to the above comparison, the middleware systems can be compared on the basis of resource management, data management, communication methods and security. Globus and Legion provide broker services whereas UNICORE and Alchemi do not provide brokering of user requests. In the case of Alchemi, Gridbus broker fulfills the role of grid brokering. The Globus communications module is based on the Nexus communication library and UNICORE's communication methods are based on AJO model and does not support synchronous message passing. Legion supports a variation of RMI for communication through LOIDs. UNICORE security is based on the Secure Socket Layer (SSL) protocol and X.509V3 type certificates. Globus provides security services through GSI, which is again based on SSL and X.509 certificates. In Gridbus's Alchemi middleware, security services are leveraged from the powerful toolset in Microsoft .NET framework. Other Gridbus components extensively leverage security services provided by Globus GSI.

## 9. Summary

There has been considerable research aiming at the development of Grid middleware systems and most of these systems have been successfully applied in other Grid related projects and applications. The Globus toolkit is one of the most widely used low-level Grid middleware today. It provides key services such as resource access, data management and security infrastructure. UNICORE is a Java-based grid computing system which is being used in projects including EUROGRID and GRIP. Gridbus toolkit extensively leverages related software technologies and provides an abstraction layer to hide idiosyncrasies of heterogeneous resources and low-level middleware technologies from application developers. It focuses on realization of utility computing model scaling from clusters to grids and to peer-to-peer computing systems. Legion is an object-based middleware system implemented in projects like Boeing's R&D project.

The comparison of grid middleware systems has shown that there is considerable overlap between their functionality. The main difference has been found to be in the architecture or implementation model. The examination of UNICORE's features shows that it does not provide brokerage services that manage or schedule computations across global resources. This shortcoming motivated us to port Gridbus resource broker to the UNICORE environment.

## Acknowledgement

We would like to thank Sven van den Berghe from the Fujitsu Laboratories of Europe for his valuable technical guidance during the creation of Gridbus broker adaptor for Unicore environment.